\documentclass[aps,prl,reprint,superscriptaddress,nofootinbib,floatfix,amsmath,amssymb]{revtex4-2}

\makeatletter
\let\frontmatter@footnote\footnote
\makeatother

\usepackage{graphicx}
\usepackage{dcolumn}
\usepackage{bm}
\usepackage{xurl} 
\usepackage{hyperref}

\begin{document}


\title{Phase Transitions in Unsupervised Feature Selection}

\author{Jonathan Fiorentino\textsuperscript{*,$\dagger$}\footnotemark[1]\footnotemark[2]}
\affiliation{Center for Life Nano- and Neuro-Science, RNA Systems Biology Lab, Fondazione Istituto Italiano di Tecnologia (IIT), 00161 Rome, Italy}

\author{Michele Monti\textsuperscript{*}\footnotemark[1]}
\affiliation{Center for Life Nano- and Neuro-Science, RNA Systems Biology Lab, Fondazione Istituto Italiano di Tecnologia (IIT), 00161 Rome, Italy}

\author{Dimitrios Miltiadis-Vrachnos}
\affiliation{Center for Life Nano- and Neuro-Science, RNA Systems Biology Lab, Fondazione Istituto Italiano di Tecnologia (IIT), 00161 Rome, Italy}
\affiliation{Department of Biology and Biotechnologies, Sapienza Università di Roma, Piazzale Aldo Moro 5, 00185 Rome, Italy}

\author{Vittorio Del Tatto}
\affiliation{Scuola Internazionale Superiore di Studi Avanzati (SISSA), Via Bonomea 265, 34136 Trieste, Italy}

\author{Alessandro Laio}
\affiliation{Scuola Internazionale Superiore di Studi Avanzati (SISSA), Via Bonomea 265, 34136 Trieste, Italy}
\affiliation{International Centre for Theoretical Physics (ICTP), Strada Costiera 11, 34151 Trieste, Italy}

\author{Gian Gaetano Tartaglia\textsuperscript{$\dagger$}\footnotemark[2]}
\affiliation{Center for Life Nano- and Neuro-Science, RNA Systems Biology Lab, Fondazione Istituto Italiano di Tecnologia (IIT), 00161 Rome, Italy}
\affiliation{Centre for Human Technologies (CHT), RNA Systems Biology Lab, Fondazione Istituto Italiano di Tecnologia (IIT), 16152 Genova, Italy}


\begin{abstract}
Identifying minimal and informative feature sets is a central challenge in data analysis, particularly when few data points are available. Here we present a theoretical analysis of an unsupervised feature selection pipeline based on the Differentiable Information Imbalance (DII). We consider the specific case of structural and physico-chemical features describing a set of proteins. We show that if one considers the features as coordinates of a (hypothetical) statistical physics model, this model undergoes a phase transition as a function of the  number of retained features. For physico-chemical descriptors, this transition is between a glass-like phase when the features are few and a liquid-like phase. The glass-like phase exhibits bimodal order-parameter distributions and Binder cumulant minima. In contrast, for structural descriptors the transition is less sharp. Remarkably, for physico-chemical descriptors the critical number of features identified from the DII coincides with the saturation of downstream binary classification performance. These results provide a principled, unsupervised criterion for minimal feature sets in protein classification and reveal distinct mechanisms of criticality across different feature types.

\end{abstract}

\maketitle
\renewcommand{\thefootnote}{\fnsymbol{footnote}}
\footnotetext[1]{These authors contributed equally to this work.}
\footnotetext[2]{Corresponding authors: 
\href{mailto:jonathan.fiorentino@iit.it}{\nolinkurl{jonathan.fiorentino@iit.it}}; 
\href{mailto:gian.tartaglia@iit.it}{\nolinkurl{gian.tartaglia@iit.it}}.}
\renewcommand{\thefootnote}{\arabic{footnote}}

\section{\label{sec:level1}Introduction}

Protein classification relies on high-dimensional feature representations derived from sequence and structure. These representations are often strongly correlated, or redundant, or affected by strong noise, making feature selection essential. While supervised methods can identify discriminative features, they require labels and tend to overfit in the low-data regime~\cite{kuncheva18}. An alternative is to perform feature selection by exploiting the intrinsic geometry of the data in an unsupervised manner.

The latter approach can be naturally implemented through the Information Imbalance (II), a general information-theoretic quantity that compares the relative information content of distance measures defined on the same space~\cite{glielmo22}. Among its applications, this measure has been employed to extract informative features from clinical data sets~\cite{wild24}, evaluate the quality of chemical descriptors~\cite{donkor23}, and infer the presence of causal links from time series data ~\cite{deltatto24,allione25,deltatto25,salvagnin25}. Its differentiable formulation, the Differentiable Information Imbalance (DII), allows optimizing a feature space by gradient methods~\cite{wild25}. When the reference and input features coincide, the DII can be used in a fully unsupervised setting, and its value provides a natural order parameter measuring how much of the information conveyed by all the features is retained by a subset of features (Supplemental Material Sec. I).

Here we apply this approach to sets of features describing protein datasets. In this framework, the DII plays the role of an order parameter, while the number of non-zero features acts as a control parameter. This  allows using tools from  statistical physics of disordered systems to interpret the outcome of a feature selection pipeline. In all the feature sets we considered, we observe a transition between a low-information phase when the features are few and a high-information phase as the number of features increases. We found that the nature of this transition depends on the set of features considered: for physico-chemical descriptors, the transition is akin to a glass-to-liquid transition, whereas structural descriptors display a gradual crossover without a clear critical point.  We perform an analysis aimed at understanding this difference. Notably, physico-chemical descriptors are correlated in blocks, and reducing this correlation structure suppresses the transition. In contrast, for structural descriptors the inter-dependencies are weaker. Moreover, features with high variance are selected first, and increasing the number of non-zero features leads to a gradual emergence of the high-information phase, consistent with the diffuse correlation structure of these descriptors. For physico-chemical features, the critical point extracted from this purely unsupervised analysis coincides with the saturation of classification performance in a binary task discriminating a specific functional protein class from a negative set. Our work establishes a direct link between the statistical properties of the feature space, criticality, and generalization in protein classification.

\section{\label{sec:level2}Results}
\textbf{Criticality of the Differentiable Information Imbalance} ---
We focus on four human protein datasets of liquid–liquid phase separating (LLPS), RNA-binding proteins (RBPs), membrane and enzyme proteins. To carry out a downstream binary classification after the unsupervised feature selection, each dataset was balanced with negative examples belonging to other functional classes~\cite{monti25,vandelli23,fiorentino24,fiorentino25,mohammed15,peng22}. Proteins are described either by sequence-derived physico-chemical features or by structure-based descriptors computed from AlphaFold-predicted structures (Supplemental Material Sec. II)~\cite{klus14,jumper21,miotto19,miotto22}. In the unsupervised setting, the full feature space is used as the target for computing the DII~\cite{glielmo22dadapy}.

\begin{figure}[t]
\includegraphics[width=\columnwidth]{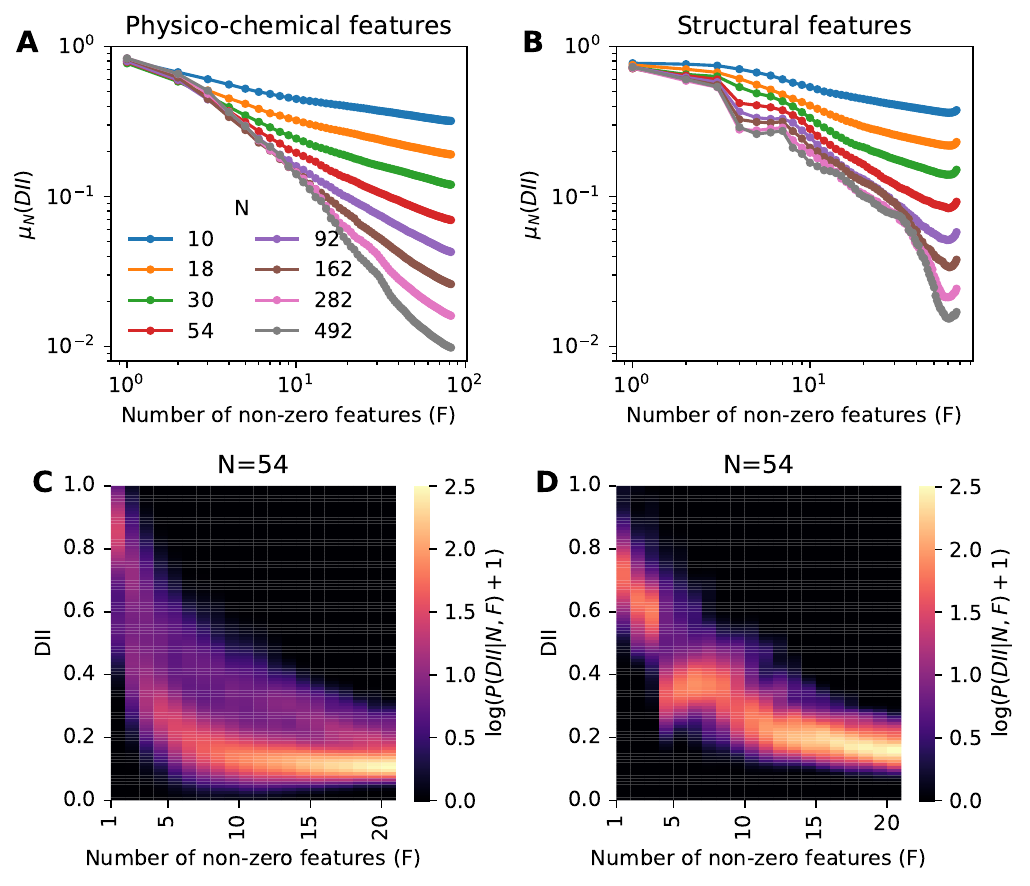}
\caption{\label{fig:1}Criticality in the Differentiable Information Imbalance during feature elimination. (A,B) Average DII versus the number of non-zero features $F$ for the LLPS dataset, for physico-chemical (A) and structural (B) features. (C,D) Heatmaps of the log-transformed probability density of the DII $P(DII|N,F)$ for the LLPS dataset, for $N=54$, for physico-chemical (C) and structural (D) features. DII values are computed on independent test random subsamples using the trained optimal weights.}
\end{figure}

Physico-chemical and structural feature sets differ markedly in their statistical properties. Physico-chemical features have few zero values, have approximately Gaussian distributions, and are strongly redundant, forming well-defined blocks of correlated variables. In contrast, structural features have a larger fraction of zero values, display heterogeneous distributional shapes, and weaker, less structured correlations (Supplemental Material Sec. III, Fig. S1). To investigate how the relevance of features changes in low-data regimes, we performed a large-scale downsampling analysis. We generated multiple balanced subsets of proteins at varying sample sizes $N$ and performed backward feature elimination using the DII~\cite{wild25} (Supplemental Material Sec. IV), which yields a DII value for each number of non-zero features $F$. Next, for each training random sample we computed the DII on an equally sized independent test set, using the weights optimized on the training set. Figure~\ref{fig:1}A,B shows the average DII as a function of the number of non-zero features $F$ for physico-chemical and structural descriptors, within the LLPS dataset. In both cases the DII decreases as features are added, but with markedly different phenomenology. For physico-chemical features the decay is smooth and monotonic, while for structural features it is more irregular, reflecting their heterogeneous  distributions (Supplemental Material Sec. IV, Fig. S2).

In order to better characterise the landscape of the DII as a function of the number of features $F$, for each value of $F$ we select at random $N$ data points and we minimise the DII for that specific dataset. This allows obtaining multiple values of the DII corresponding to different random subsets of identical size $N$. We used these values to compute the empirical distribution $P(DII|N,F)$, which is shown in Fig.~\ref{fig:1}C,D for $N=54$. For the physico-chemical features $P(DII|N,F)$ exhibits a bimodal distribution at small F, which becomes unimodal as F increases—an archetypal signature of a glassy landscape with competing minima, resembling the order parameter distributions seen in the Biroli–Mézard lattice glass model~\cite{hukushima10,biroli01lattice}. This bimodality indicates that distinct subsets of features yield comparable DII values at fixed $F$, revealing a degeneracy of near-optimal solutions. For structural features, instead, $P(DII|N,F)$ displays a single peak for all values of $F$, indicating a less frustrated optimization landscape. As shown in Supplemental Material Sec. IV, Fig. S3, the same qualitative behaviour is well reproduced across all datasets and for other values of $N$.

\textbf{Finite-size scaling and critical feature number} ---
To characterize the transition quantitatively, we compute the Binder cumulant $U(F)$ of the DII distribution as a function of $F$ (Supplemental Material Sec. IV, Fig. S4)~\cite{binder81}. 
\begin{figure*}[tb]
\includegraphics[width=\textwidth]{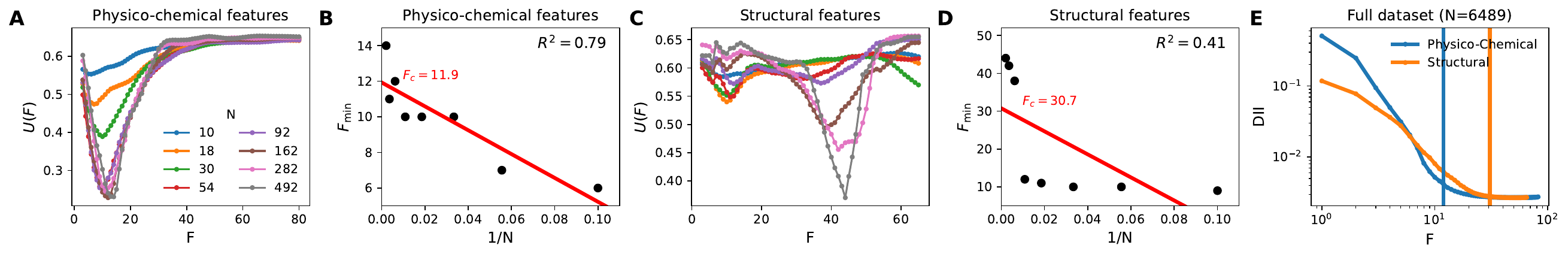}
\caption{\label{fig:2} Binder cumulant analysis reveals a glass-like phase transition for physico-chemical features. (A,C) Binder cumulant $U(F)$ as a function of the number of non-zero features $F$ for physico-chemical (A) and structural (C) descriptors, for the LLPS dataset. (B,D) Extrapolation of $F_{min}$ (position of Binder minimum) versus $1/N$, indicating the critical feature number $F_c$ for physico-chemical (B) and structural features (D). (E) DII as a function of the number of non-zero features, computed on the full LLPS dataset ($N=6489$ proteins), for physico-chemical (blue) and structural (orange) features. The blue and orange lines indicate the values of $F_c$ for the two sets of features.}
\end{figure*}
As shown in Fig.~\ref{fig:2}A, physico-chemical features display a pronounced minimum of the Binder cumulant for small values of $F$. The dependence on dataset size $N$ of the position of this minimum, shifting towards larger $F$ for larger samples, is a finite-size precursor of a phase transition~\cite{binder81}. Extrapolating the position of this minimum to infinite sample size, we define the critical number of features $F_c$ as the large-$N$ limit of the Binder cumulant minimum for physico-chemical descriptors (Fig.~\ref{fig:2}B)~\cite{hukushima10,biroli01lattice}. Structural features show a shallower minimum of $U(F)$ (Fig.~\ref{fig:2}C) and display substantially weaker criticality (Fig.~\ref{fig:2}D). This is confirmed on the full datasets (Fig.~\ref{fig:2}E): for physico-chemical features, the DII rises sharply when $F < F_c$, indicating that the remaining features quickly lose informativeness. In contrast, for structural features the DII decreases more gradually as a function of $F$, showing that informativeness about the full space is more homogeneously distributed across features in this set. This makes feature selection more challenging for these descriptors.

\begin{figure}[b]
\includegraphics[width=\columnwidth]{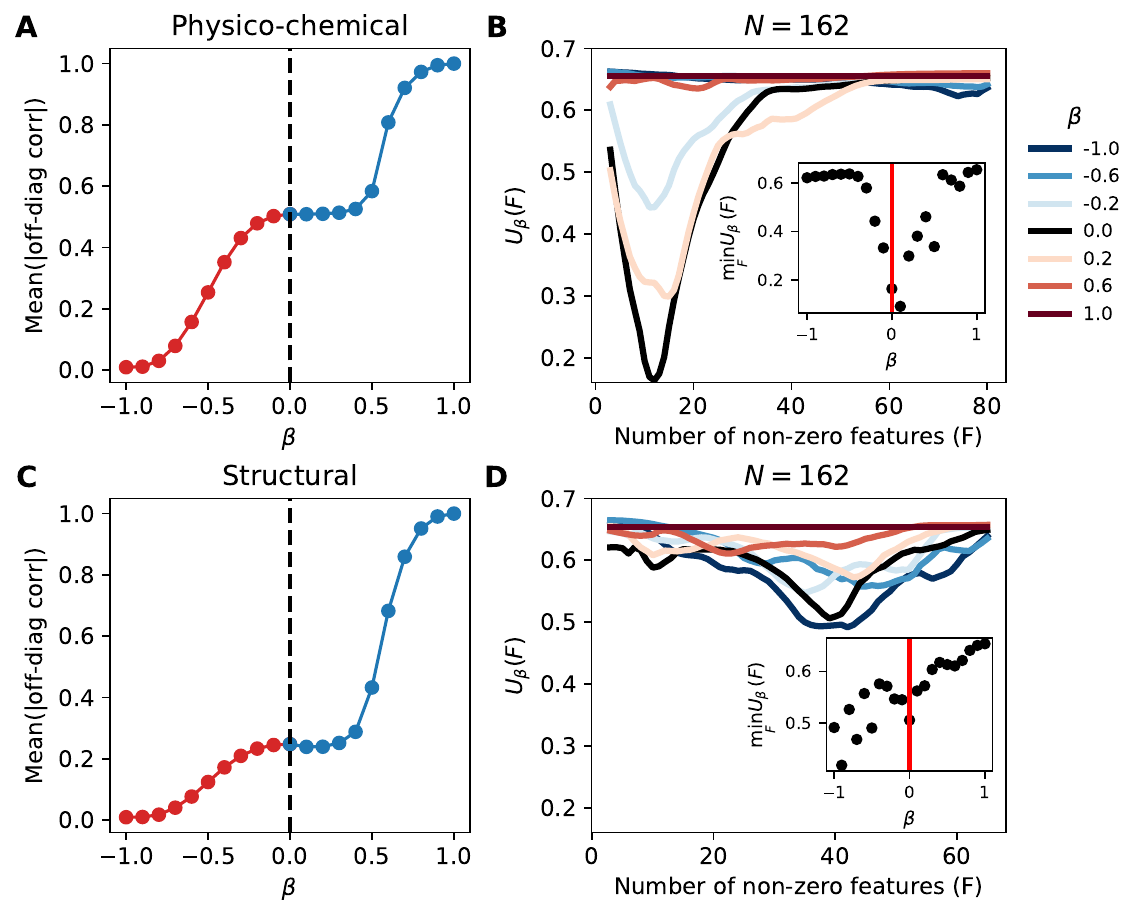}
\caption{\label{fig:3}Feature-set dependent origin of the phase transition. (A,C) Mean absolute off-diagonal correlation of the physico-chemical (A) and structural (C) feature matrix as a function of the tuning parameter $\beta$. (B,D) Binder cumulant curves $U(F)$ for varying $\beta$ values at fixed $N=162$, for physico-chemical (B) and structural (D) features. The insets show the minimum value of $U(F)$ as a function of $\beta$ at $N=162$. The red line indicates $\beta=0$.}
\end{figure}

\textbf{Correlation-driven criticality vs variance-driven crossover} --- To uncover the mechanisms underlying the observed transitions, we developed a model in which  the correlation structure of the feature sets can be tuned using a parameter $\beta$. For $\beta=0$ the correlations are unchanged. For $\beta<0$ the correlation is reduced,  while for $\beta>0$ the correlation is increased (Fig.~\ref{fig:3}A-C, Supplemental Material Sec. V, Fig. S5).
For physico-chemical features (Fig.~\ref{fig:3}B), the glass-like phase transition emerges only within an intermediate correlation regime around the empirical value ($\beta=0$). Suppressing or excessively amplifying correlations eliminates the Binder minimum, suggesting that the block structure in the correlations is important to observe such a transition. The block structure indicates that features behave as interacting degrees of freedom, whose effective constraints generate frustration and multiple near-degenerate solutions, in close analogy to what happens in  lattice glass models~\cite{biroli01lattice}. In contrast, structural features (Fig.~\ref{fig:3}D) display a different pattern. At $\beta>0$, increasing correlations diminishes the minimum and eventually suppresses the transition, as observed for physico-chemical features. Instead, at $\beta<0$, where correlations are suppressed, the Binder minimum becomes sharper, although much less pronounced than for physico-chemical features. This behavior indicates that for structural features, the phase transition is not driven by the correlation structure but rather by heterogeneity in feature variances (Fig. S6A), as confirmed by the disappearance of the phase transition when feature variances are made homogeneous in absence of correlations (Fig. S6B-C). These results highlight that the mechanisms determining the information content as a function of the number of features are feature-set dependent: physico-chemical features undergo a correlation-driven, glass-like transition, whereas structural features, which are weakly correlated, exhibit a variance-driven transition that is further sharpened by reducing correlations.

\begin{figure}[t]
\includegraphics[width=\columnwidth]{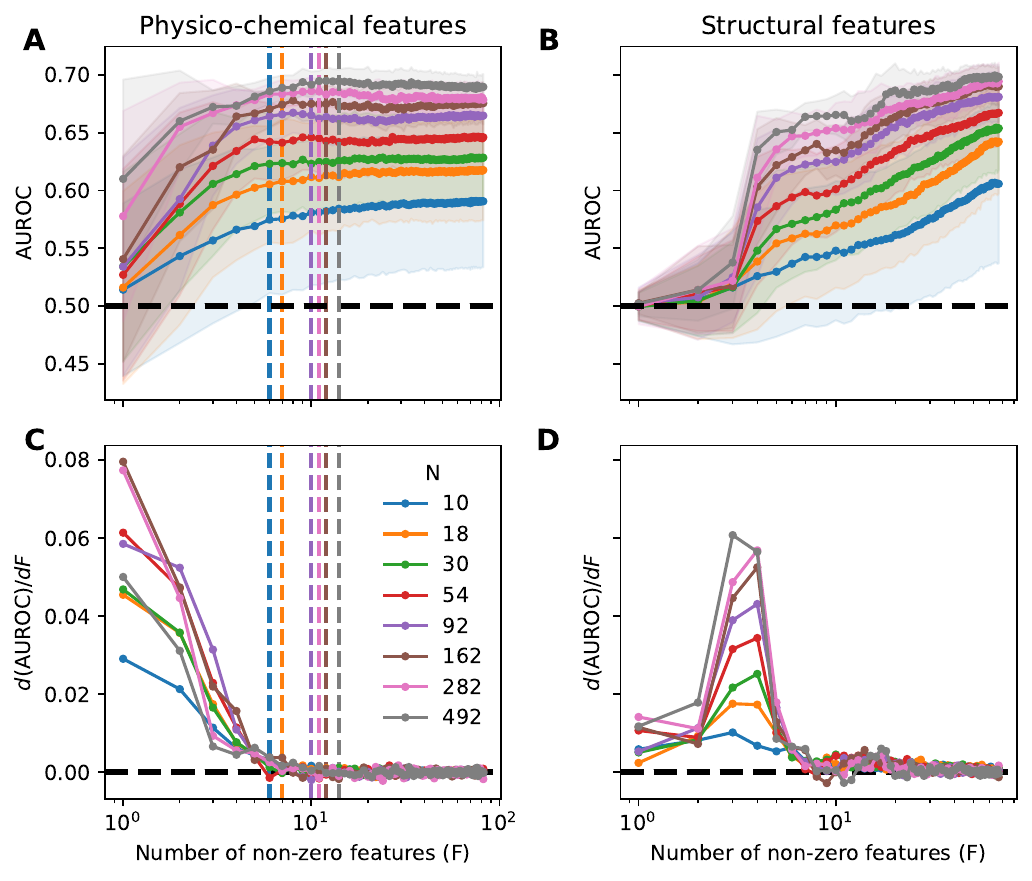}
\caption{\label{fig:4}Binary classification performance saturates at the critical point for physico-chemical features. (A,B) Area under the receiver operating characteristic (AUROC) as a function of the number of non-zero features $F$ for the LLPS dataset, using physico-chemical (A) and structural (B) features. Solid lines show the mean over subsamples, shadowed areas the standard deviation. Colored dashed vertical lines indicate the values of $F_{min}(N)$ extracted from the Binder cumulant analysis. The black dashed horizontal line denotes the performance of a random classifier. (C,D) Numerical derivative of the AUROC with respect to $F$ for physico-chemical (C) and structural (D) features. AUROC is computed on a held out test set (Supplemental Material Secs. I and VIII). The black dashed horizontal line shows where the derivative is zero.}
\end{figure}

\textbf{Alignment between unsupervised criticality and supervised performance} ---
Finally, we relate the critical number of features $F_c$ to binary classification performance. For each number of features $F$ selected by the DII we trained a  multilayer perceptron to classify LLPS proteins (Supplemental Material Sec. VI, Fig. S7). As shown in Fig.~\ref{fig:4}A-C, for physico-chemical features the classification performance, quantified by the area under the receiver operating characteristic (AUROC), increases with $F$ and saturates at the critical feature number $F_c$, with the saturation becoming progressively sharper as the sample size $N$ increases. Beyond this point, adding further features yields negligible improvement. In contrast, for structural features (Fig.~\ref{fig:4}B-D) the AUROC does not exhibit a clear plateau as $F$ increases, but instead grows smoothly over the entire range.  This behavior indicates the absence of a sharply defined minimal feature subset capable of fully capturing the information content of the structural representation.

\section{\label{sec:level2}Discussion}

In this work we have shown that unsupervised feature selection based on the Differentiable Information Imbalance exhibits a glass-like phase transition as the number of retained features is varied, depending on the structure of the feature space. Treating the DII as an order parameter and the number of non-zero features as a control parameter, we observe hallmark signatures of criticality—including bimodal order-parameter distributions and minima in the Binder cumulant—closely analogous to lattice glass transitions in disordered systems.  Although no explicit Hamiltonian governs feature selection, these signatures allow us to operationally define a phase-transition–like behavior. 

The emergence of this transition is feature-set dependent. For sequence-derived physico-chemical descriptors, which are characterized by strong block correlations and redundancy, the DII landscape becomes rugged below a critical number of features, with many distinct subsets yielding nearly equivalent values of the order parameter. This phenomenology is reminiscent of glassy phases in spin systems, where competing constraints generate a proliferation of metastable states. In contrast, structural features show a qualitatively different behavior. While a crossover in the DII can still be identified, its signatures are substantially weaker: Binder cumulant minima are shallow and the probability density of the DII is unimodal. Importantly, this difference cannot be explained by the difference in  correlations. Structural descriptors are intrinsically more heterogeneous, with broader and more uneven feature variances, and this heterogeneity drives a crossover rather than a phase transition.

By systematically perturbing the feature space, we separate the mechanisms underlying the transitions. For physico-chemical features, suppressing or excessively amplifying correlations eliminates the Binder cumulant minimum, showing that block correlations and redundancy drive the criticality. In contrast, for structural features, disrupting correlations makes the crossover slightly sharper, but no bimodality in the probability distribution of the DII appears. Homogenizing feature variances in an uncorrelated setting suppresses the crossover, indicating that heterogeneity, rather than correlations, determines the feature selection landscape for structural features.

A central result of this work is the direct connection between the critical point of the DII, when it exists,  and binary classification performance. We find that, for physico-chemical features, the critical number of features identified from the DII coincides with the saturation of binary classification performance. Remarkably, this optimal compression scale is identified without access to labels, indicating that the geometry of the feature space alone encodes the limits of predictive performance.

More broadly, our results suggest that feature selection in high-dimensional data can be understood using tools from  the  statistical physics of disordered systems. In this view, informative features act as interacting degrees of freedom subject to competing constraints, and generalization emerges at the edge of a glassy phase. 

This perspective resonates with an information-theoretic interpretation in which criticality arises as a hallmark of maximally informative representations, obtained at an optimal trade-off between resolution and relevance, where power-law statistics and Zipf-like behavior naturally emerge without parameter fine tuning~\cite{cubero19,marsili13sampling}. While that approach links criticality to the statistics of observed state frequencies, our results establish a connection between information optimality and the geometry of correlated feature spaces, revealing a glass-like transition driven by redundancy and frustration.

In perspective our results may open the door to a theoretical characterization of feature selection in terms of replica symmetry breaking and cavity-based approaches~\cite{biroli01lattice,mezard03}, and points to criticality as a general principle for determining optimal representations in complex data.

The code used to run the analyses, as well as the raw data, is available at \url{https://github.com/tartaglialabIIT/DIIPhaseTransition.git} and \url{https://doi.org/10.5281/zenodo.18223323}.

\section{\label{sec:level3}Acknowledgments}

The research leading to these results have been supported through ERC [ASTRA\_855923 (to G.G.T.), H2020 Projects IASIS\_727658 and INFORE\_825080 and IVBM4PAP\_101098989 (to G.G.T.)] and National Center for Gene Therapy and Drug based on RNA Technology (CN00000041), financed by NextGenerationEU PNRR MUR - M4C2 - Action 1.4- Call “Potenziamento strutture di ricerca e di campioni nazionali di R\&S” (CUP J33C22001130001) (to G.G.T.).

\bibliography{biblio}

\end{document}


\title{Supplemental Material for ``Phase Transitions in Unsupervised Feature Selection''}
\maketitle

\section*{I. Differentiable Information Imbalance (DII)}
The Differentiable Information Imbalance quantifies how much an input distance space is predictive of a reference distance space, by measuring how faithfully similarity relationships in the former are reproduced in the latter. Following ref.~\cite{wild25}, we consider two different feature representations of the same points, which we denote by $A=\{x_i\}_i$ and $B=\{y_i\}_i$, with $i=1,...,N$. In the reference space $B$, distances are defined as standard Euclidean distances,

\begin{equation}
d_{ij}^B = \| y_i - y_j \|, 
\tag{S1}
\end{equation}
	
where $\| . \|$ is the Euclidean norm. Neighbourhood relationships are encoded through distance ranks ${r_{ij}^B}$, where $r_{ij}^B$ denotes the neighbour order of $j$ with respect to $i$ according to distance $d^B$. For example, $r_{ij}^B=2$ if $j$ is the second nearest neighbour of $i$. In the input space $A$, in contrast, similarities are measured using weighted Euclidean distances,
\begin{equation}
	d_{ij}^A = \| w \odot (x_i - x_j) \|,
\tag{S2}
\end{equation}

where $w$ is a vector of feature-specific scaling parameters and $\odot$ denotes the element-wise product. Such weights are introduced in order to optimise distance $d^A$ through the  minimization of the following loss function, which defines the DII from $A$ to $B$:
\begin{equation}
	DII(A \rightarrow B) = \frac{2}{N^2}\sum_{ij(i\neq j)} c_{ij}^A(w) r_{ij}^B.
\tag{S3}
\end{equation}
The coefficients $c_{ij}^A(w)$ are defined via a softmax over distances in space $A$,
\begin{equation}
	c_{ij}^A(w) = \frac{exp(-d_{ij}^A/\lambda)}{\sum_{k(\neq i)} exp(-d_{ik}^A/\lambda)},
\tag{S4}
\end{equation}
where the parameter $\lambda$ sets a scale of ``closeness'' in space A, which can be fixed, for example, using the average distances of first and second nearest neighbours~\cite{wild25}.

For each point $i$, the coefficients $c_{ij}^A(w)$ are significantly nonzero only for points $j$ that are close to $i$ in space $A$. Therefore, the DII is minimised when points that are close in space $A$ are also close in space $B$, namely, when the ranks $r_{ij}^B$ selected by the coefficients $c_{ij}^A(w)$ are as small as possible. By minimizing $DII(A \rightarrow B)$ via gradient-descent optimization, one learns the weighted distance $d^A$ that best reproduces the neighborhood structure of the reference space $B$.

The DII framework can be employed for feature selection through several strategies, such as backward greedy elimination. In this approach, the DII optimization is repeated iteratively while removing one feature at a time from space $A$, and the resulting optimal DII values are compared. The optimization associated with the largest DII identifies the least informative feature, which is then removed. Repeating this procedure until a single feature remains produces a natural ranking of feature importance.
Although the procedure described above defines a supervised protocol, the same principles apply when the input and reference spaces coincide ($A=B$). In this case, feature selection identifies the variables that contribute most to the neighbourhood structure of the full space, yielding a fully unsupervised feature ranking.

\section*{II. Datasets and protein feature representation}
We analyzed four balanced human protein datasets corresponding to distinct functional classes: liquid–liquid phase separation (LLPS), RNA-binding proteins (RBP), membrane proteins, and enzymes. Within each dataset, positive and negative sets were obtained or curated from publicly available sources (catGRANULE 2.0~\cite{monti25}, PRALINE~\cite{vandelli23}, scRAPID~\cite{fiorentino24,fiorentino25}, RBP TS-TL~\cite{peng22}, UniProt, ECemble~\cite{mohammed15}). Except for the LLPS dataset, which was already filtered, balanced, and split into training and test sets, each dataset was filtered at $50\%$ sequence identity using CD-HIT and restricted to proteins with AlphaFold2-predicted structures~\cite{jumper21}. Each dataset was then balanced by random down-sampling of the negative class, when necessary, and split into training and test subsets using an $80/20$ ratio. Table~\ref{tab:S1} summarizes the sizes of the resulting datasets.

\begin{table}
\caption{Number of proteins in the training and test sets for each functional class (LLPS, RNA-binding proteins, membrane proteins, and enzymes). Each dataset includes two balanced sets: positive examples (proteins belonging to the reference class) and negative examples (proteins belonging to any other functional class).}
\label{tab:S1}
\begin{tabular}{lcc}
\hline
Dataset & Train & Test \\
\hline
LLPS      & 6489 & 2768 \\
RBP       & 4281 & 1071 \\
MEMBRANE  & 6181 & 1547 \\
ENZYME    & 1431 & 359  \\
\end{tabular}
\end{table}

Each protein was represented by two complementary feature sets:
\begin{itemize}
    \item  Physico-chemical features ($N=82$): Sequence-derived descriptors from curated amino acid scales capturing hydrophobicity, aggregation, disorder, nucleic acid binding, and secondary-structure propensities~\cite{monti25,klus14}. Each feature is computed as the average of residue-scale values along the sequence. Physico-chemical features range in $[0,1]$, except for charge.
    \item Structural features ($N=67$): Computed from AlphaFold2 models, including global geometric descriptors (radius of gyration, surface area, contact number), disorder and secondary-structure composition, and graph-theoretical features derived from residue–residue interaction networks~\cite{monti25,miotto19}. Energetic features describing Coulombic and van der Waals interactions were estimated using the Thermometer framework~\cite{miotto22}. All features were normalized by Min–Max scaling fitted on the human proteome to ensure comparability with physico-chemical features.
\end{itemize}

\section*{III. Statistical Properties of the Feature Sets}
\begin{figure}[b]
  \centering
  \includegraphics[width=0.6\textwidth]{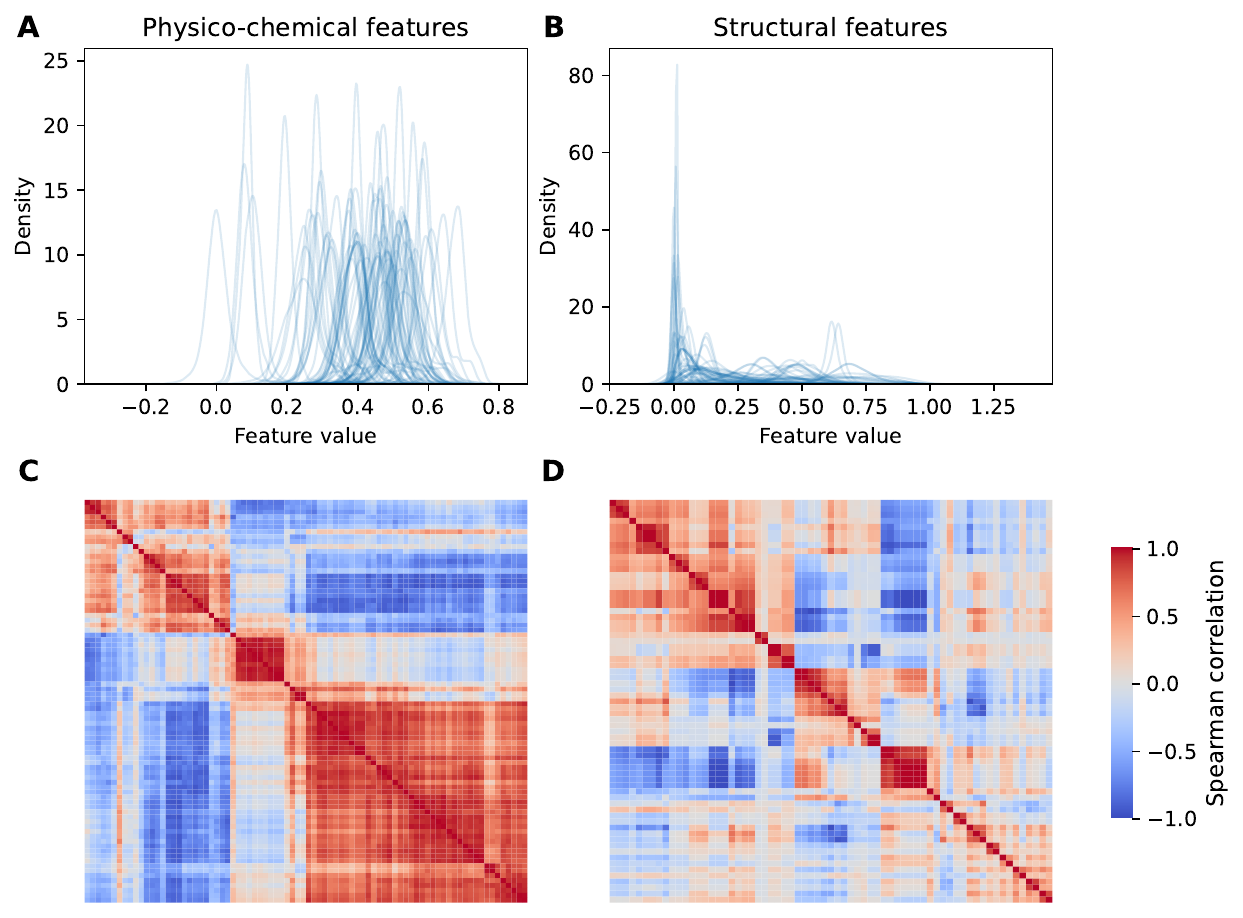}
  \caption{(A-B) Probability densities of distinct physico-chemical and structural features, for the LLPS dataset. (C-D) Spearman's correlation cluster maps for the LLPS dataset, for physico-chemical and structural features.}
  \label{fig:S1}
\end{figure}

We characterized the statistical structure of the physico-chemical and structural feature representations by analyzing the distributions of feature values and pairwise correlations across all datasets. Physico-chemical features exhibit near-Gaussian value distributions centered around $0.5$ (Fig.~\ref{fig:S1}A) and display pronounced redundancy, organized into strongly correlated blocks (Fig.~~\ref{fig:S1}C). In contrast, structural features are markedly sparser and more heterogeneous, with broadly distributed value profiles (Fig.~~\ref{fig:S1}B) and weaker, more diffuse correlations (Fig.~~\ref{fig:S1}D). 

\section*{IV. Dataset downsampling, DII feature selection and extended results on the phase transition}

To investigate the scaling of Differentiable Information Imbalance (DII) varying the number of proteins in the datasets, we generated random subsamples of increasing size
$N = 10, 18, 30, 54, 92, 162, 282, 492$,
each balanced between positive and negative examples.
For each $N$, we produced $M$ independent subsamples depending on size ($750, 416, 250, 138, 81, 60, 40,$ and $30$, respectively) to average over sampling fluctuations. Unsupervised feature selection was performed via backward greedy elimination under the DII framework implemented in DADApy, treating the full feature set as the target distribution~\cite{wild25,glielmo22dadapy}. For each subsample, this procedure returns the DII value and associated optimal feature weights at each elimination step. This process was repeated independently for each protein dataset, for the physico-chemical and structural features. To avoid overfitting, for each training subsample, we generated an independent balanced test subsample of the same size and we computed the DII using the previously learned weights.

\begin{figure}[b]
  \includegraphics[width=\linewidth]{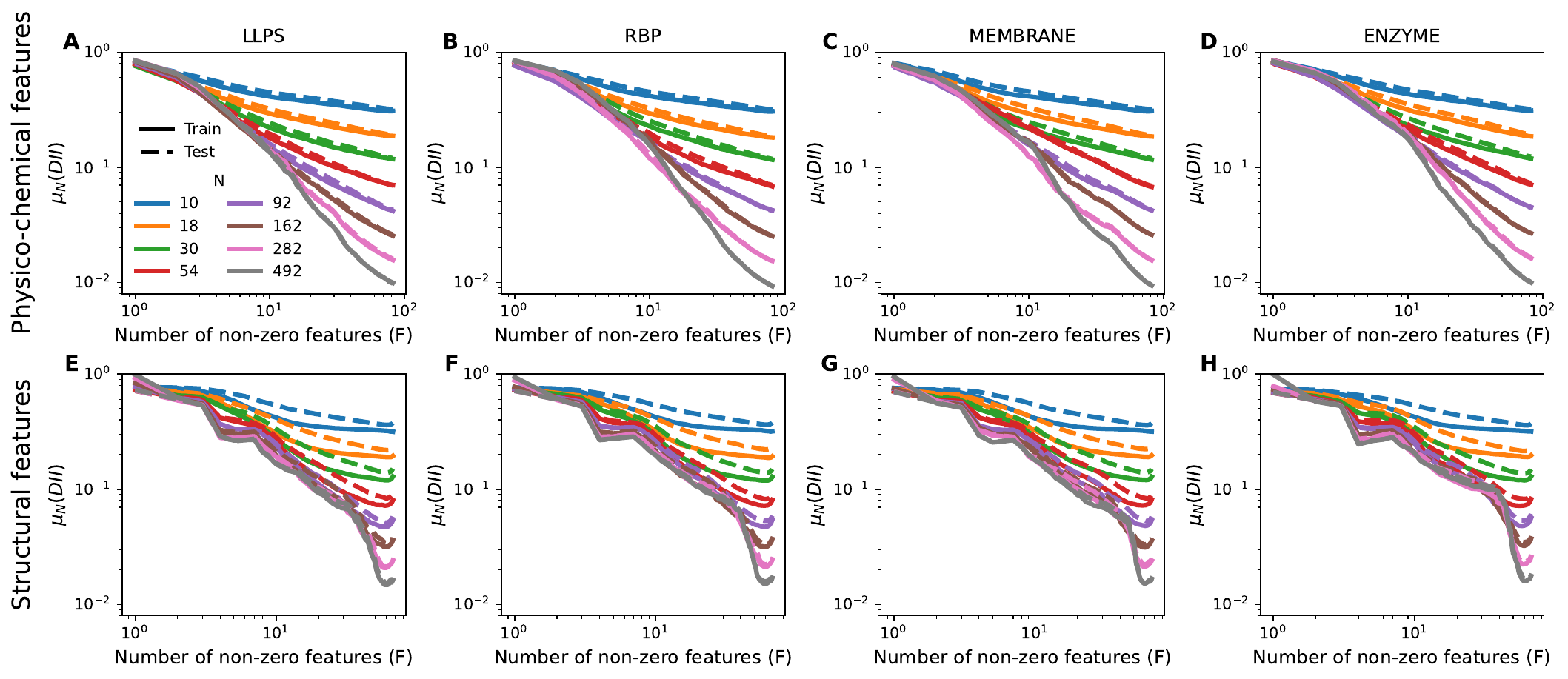}
  \caption{Average DII versus the number of non-zero features $F$ for the four protein datasets, for physico-chemical (A-D) and structural (E-H) features. Solid and dashed lines indicate training and test samples, respectively.}
  \label{fig:S2}
\end{figure}

\begin{figure}
  \includegraphics[width=\linewidth]{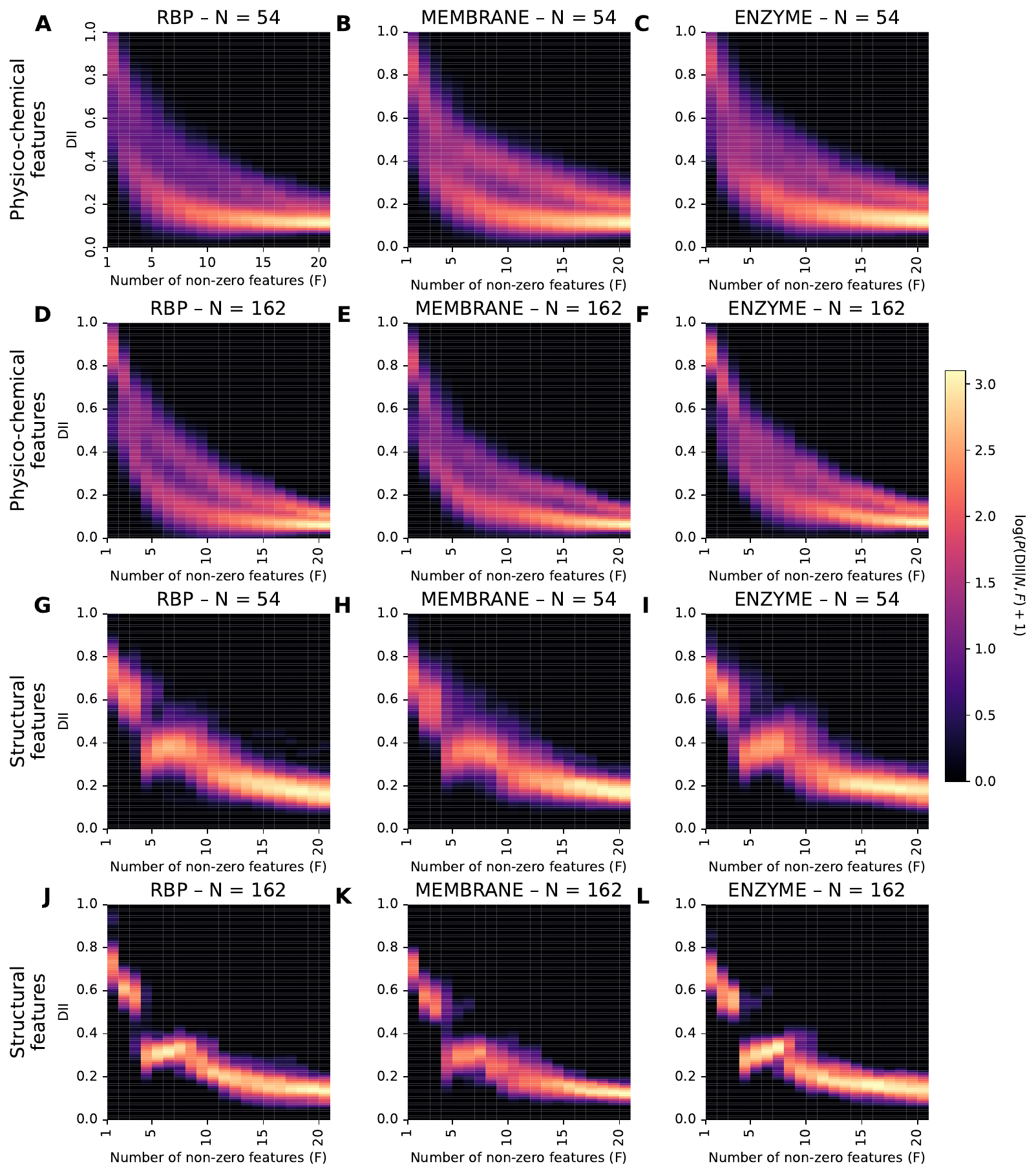}
  \caption{Heatmaps of the probability density of the DII $P(DII|N,F)$ computed over independent test subsamples of the RBP, MEMBRANE and ENZYME datasets, for $N=54$ and $N=162$, for physico-chemical (A-F) and structural (G-L) features.}
  \label{fig:S3}
\end{figure}

\begin{figure}[b]
  \includegraphics[width=0.81\textwidth]{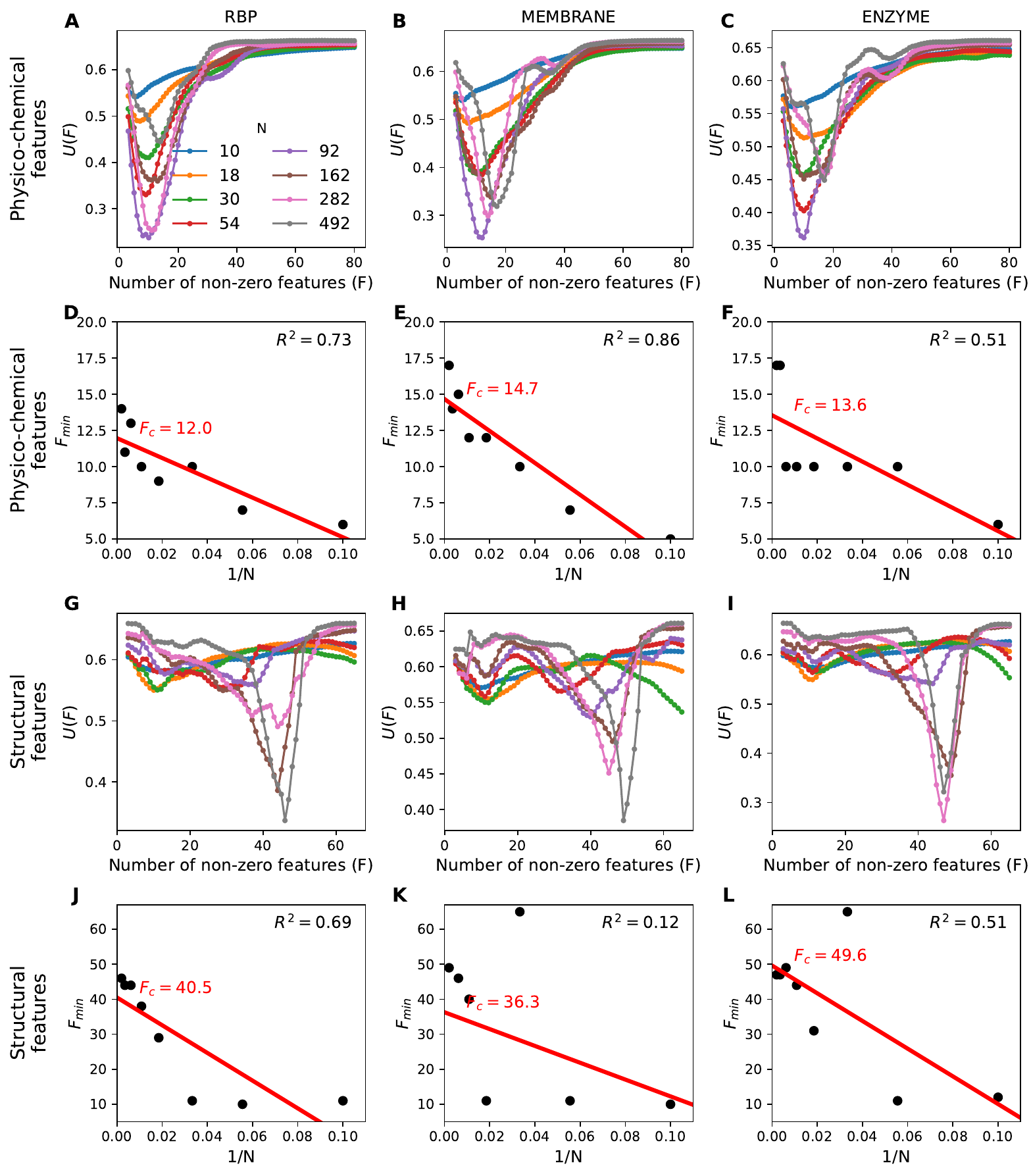}
  \caption{Binder cumulant $U(F)$ as a function of number of non-zero features $F$ for physico-chemical (A-C) and structural (G-I) descriptors, for the RBP, MEMBRANE and ENZYME datasets (columns). Extrapolation of $F_{min}$ (position of Binder minimum) versus $1/N$, indicating the critical feature number $F_c$ for physico-chemical features (D-F) and structural features (J-L).}
  \label{fig:S4}
\end{figure}

Extended analyses on the phase transition of the DII for all the protein datasets are shown in Figures \ref{fig:S2}-\ref{fig:S4}. Fig.~\ref{fig:S2} shows the average DII over the random subsamples as a function of the number of non-zero features ($F$), for different values of $N$ (number of proteins in the dataset). We observe that the difference between the average DII obtained from training and test subsamples is close to zero for the physico-chemical features and more appreciable for structural features, although in both cases it reduces increasing $N$.

Fig.~\ref{fig:S3} shows the probability density of the DII over the random subsamples, computed on the test samples, at $N=54$ and $N=162$, for the RBP, MEMBRANE and ENZYME datasets, confirming the emergence of bimodality of $P(DII|N,F)$ for physico-chemical features, as observed for the LLPS dataset.

The DII trajectories obtained for each dataset and sample size were analyzed to identify phase transitions in the feature space. The Binder cumulant $U(F)$~\cite{binder81} was computed as:
\begin{equation}
	U = 1 - \frac{\langle \Delta^4 \rangle}{3\langle \Delta^2 \rangle^2},
\tag{S5}
\end{equation}
where $\Delta$ denotes the DII value, and angular brackets indicate averaging over random subsamples. For each dataset size $N$, $U(F)$ was smoothed via a linear convolution, and the feature number $F_{min}(N)$ corresponding to the minimum of $U(F)$ was recorded. The convolution kernel has a width of 5 features; we verified that the Binder cumulant minimum is also present in the unsmoothed data, albeit with increased statistical noise. The critical number of features $F_c$ was then obtained by linear extrapolation of $F_{min}$ vs $1/N$.

Fig.~\ref{fig:S4} shows the Binder cumulant of the DII distribution, as a function of the number of non-zero features, at different values of $N$, and $F_{min}$ (number of features at which the Binder cumulant is minimum, at a given $N$), as a function of $1/N$ for the RBP, MEMBRANE and ENZYME datasets, for the physico-chemical and structural features.

For physico-chemical features, all datasets display a minimum in the Binder cumulant (Fig.~\ref{fig:S4}A-C) and a convergence of $F_{min}$ towards a finite $F_c$ as $1/N \rightarrow 0$ (Fig.~\ref{fig:S4}D-F), although $F_{min}$ values are noisier for the ENZYME dataset. This means that below $F_c$ many distinct feature subsets yield nearly identical DII values. This degeneracy indicates the emergence of a rugged DII landscape below $F_c$, consistent with a glass-like phase characterized by extensive redundancy and competing correlations among features. In contrast, structural features give rise to a much smoother evolution of the DII: the transition is less sharply defined when $N$ is small (Fig.~\ref{fig:S4}G-I), while for larger values of $N$ a clearer minimum in $U(F)$ appears. As observed for the LLPS dataset, the linear scaling of $F_{min}$ with $1/N$ is less evident than in the case of physico-chemical features (Fig.~\ref{fig:S4}J–L).

\newpage

\section*{V. Extended results on the origin of the DII phase transition}

To dissect the role of feature correlations in driving the observed DII phase transition, we constructed a model in which the correlation structure of the feature sets can be tuned via a parameter $\beta$. For $\beta < 0$, Gaussian noise is added to each feature, progressively decorrelating them, while for $\beta > 0$, the features are increasingly aligned along the first right singular vector of the centered data matrix, thereby enhancing correlations. Additionally, to explore the effect of feature heterogeneity on the transition, we introduced a second parameter $\alpha$ that controls the homogenization of feature variances: starting from their original values ($\alpha = 0$), increasing $\alpha$ gradually forces all feature variances to approach the average variance across features, while preserving the feature means. This framework allows us to systematically tune both the correlation structure and variance distribution of the feature matrix, isolating their respective contributions.

We define the original data matrix as $X \in \mathbb{R}^{N \times F}$, where $N$ is the number of proteins and $F$ is the number of features.  
For each feature $j = 1, \dots, F$, let 
\begin{equation}
\mu_j = \text{mean}(X_j), \qquad \sigma_j = \text{std}(X_j),
\tag{S7}
\end{equation}
be its mean and standard deviation, respectively.

We then define a matrix of independent Gaussian noises scaled to match the original variances:
\begin{equation}
(X_{\text{noise}})_{ij} = \sigma_j Z_{ij}, \qquad Z_{ij} \sim \mathcal{N}(0,1),
\tag{S8}
\end{equation}
where $Z_{ij}$ are i.i.d. standard normal random variables.

Next, we compute the centered version of the data matrix:
\begin{equation}
X_{\text{centered}} = X - \mu,
\tag{S9}
\end{equation}
and perform its Singular Value Decomposition (SVD). Let $\underline{v}_1$ be the first right singular vector, and define the matrix $S$ as
\begin{equation}
S = \frac{X_{\text{centered}} \underline{v}_1}{\| X_{\text{centered}} \underline{v}_1 \|}.
\tag{S10}
\end{equation}

From $S$ we construct the matrix $X_c$:
\begin{equation}
(X_c)_{ij} = \mu_j + \sigma_j S_i, \qquad \sigma_j > 0.
\tag{S11}
\end{equation}

Finally, the modified matrix at a given value of $\beta$ is defined as
\begin{equation}
X_{\text{new}} = (1 - |\beta|) X + |\beta| \big[ \Theta(\beta) X_c + \Theta(-\beta) X_{\text{noise}} \big],
\tag{S12}
\end{equation}
where $\Theta(\cdot)$ is the Heaviside step function.

In particular:
\begin{itemize}
\item For $\beta = 0$, $X_{\text{new}} = X$, and the feature correlation matrix is the original one.
\item For $\beta = -1$, $X_{\text{new}} = X_{\text{noise}}$ (Fig.~S5A).
\item For $\beta = 1$, $X_{\text{new}} = X_c$ (Fig.~S5B).
\end{itemize}

\begin{figure}
  \includegraphics[width=\linewidth]{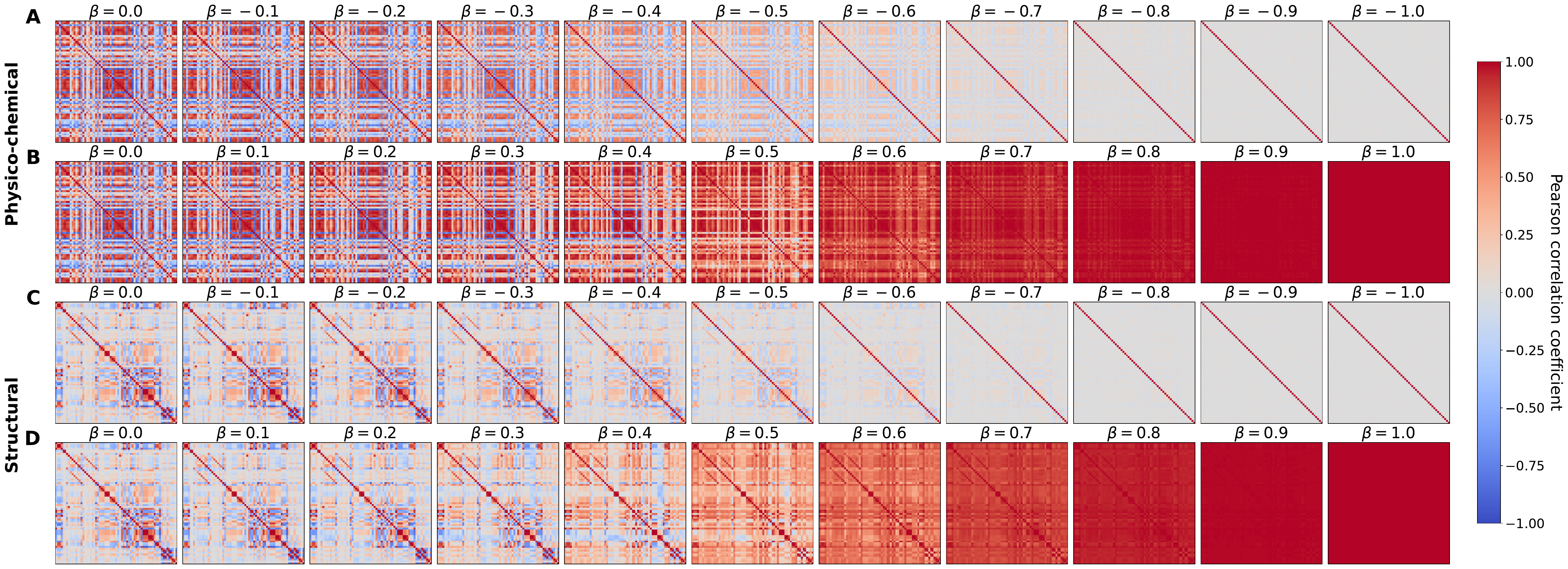}
  \caption{(A-B) Correlation matrices of the LLPS dataset, for physico-chemical features, for negative (A) and positive (B) values of $\beta$. (C-D) Correlation matrices of the LLPS dataset, for structural features, for negative (C) and positive (D) values of $\beta$.}
  \label{fig:S5}
\end{figure}

Intermediate values of $\beta$ provide partial increase or decrease of the overall feature correlation, as shown in Fig.~\ref{fig:S5}.

We note that varying $\beta$ only modifies the structure of the feature correlation matrix, without coherently changing the underlying features themselves. As a consequence, distances between proteins in the modified feature space are not preserved, and the procedure is intended to probe the role of correlation structure rather than to maintain the original geometric relationships between data points.  

\begin{figure}
  \includegraphics[width=\linewidth]{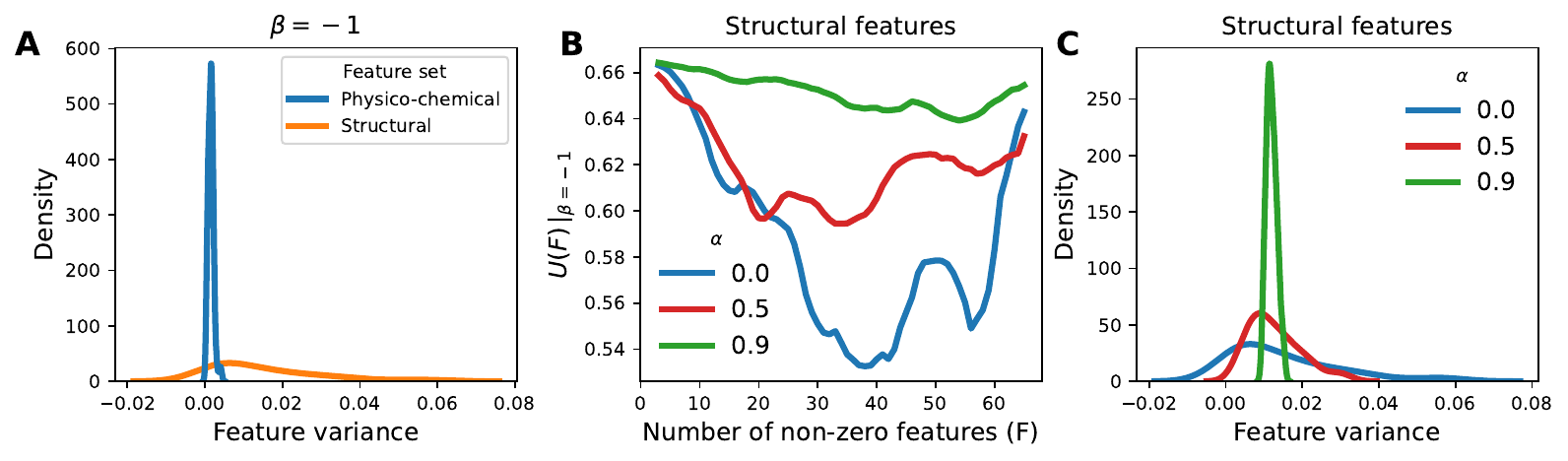}
  \caption{(A) Distribution of feature variances of physico-chemical and structural descriptors, at $\beta= -1$. (B) Binder cumulant as a function of the number of non-zero features, for structural descriptors at $\beta= -1$ and different values of $\alpha$. (C) Distribution of feature variances for structural descriptors at $\beta= -1$ and different values of $\alpha$. All panels are obtained on the LLPS dataset, $N = 162$ in (B) and (C).}
  \label{fig:S6}
\end{figure}

Next, to assess the role of feature heterogeneity in driving the transition for structural features, independently of correlations, we introduced a controlled procedure that progressively homogenizes feature variances while preserving feature means. Starting from the fully decorrelated dataset ($\beta=-1$; Fig.~\ref{fig:S6}A), each feature $x_j$ was transformed as
\begin{equation}
    x_j^{(\alpha)} = (1-\alpha)\,x_j \;+\; \left[ \alpha\,\frac{x_j-\langle x_j\rangle}{\sigma_j}\,\bar{\sigma} \;+\;\langle x_j\rangle \right],
    \tag{S13}
\end{equation}

where $\sigma_j$ is the original standard deviation of feature $j$, $\bar{\sigma}$ is the average standard deviation across all features, and $\alpha \in [0,1]$ controls the degree of variance homogenization. For $\alpha=0$ the original variance heterogeneity is preserved, while for $\alpha=1$ all features have identical variance by construction. We find that increasing $\alpha$ progressively suppresses the minimum in the Binder cumulant $U(F)$, eventually eliminating the transition when variances are fully homogenized (Fig.~\ref{fig:S6}B). In parallel, the feature-variance distributions collapse onto a narrow peak (Fig.~\ref{fig:S6}C). Conversely, at low $\alpha$, where variance heterogeneity is retained, the Binder cumulant displays a minimum, signaling a crossover.

\section*{VI. Downstream Classification and Validation}
\begin{figure}
  \centering
  \includegraphics[width=0.8\linewidth]{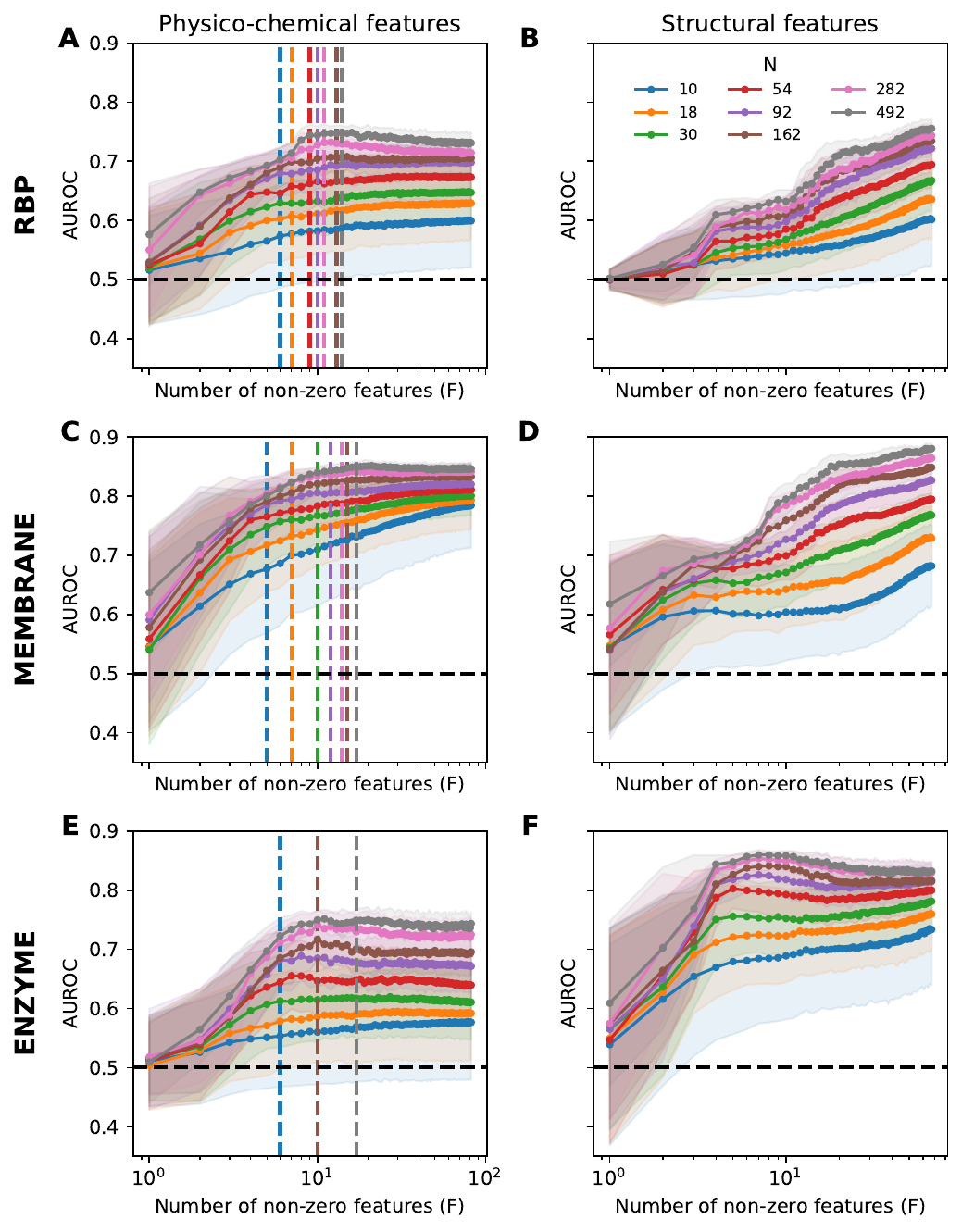}
  \caption{Area under the receiver operating characteristic (AUROC) as a function of the number of non-zero features $F$ for the RBP, MEMBRANE and ENZYME datasets, using physico-chemical (A-C-E) and structural (B-D-F) features. Colored dashed vertical lines indicate the values of $F_{min}(N)$ extracted from the Binder cumulant analysis. The black dashed horizontal line denotes the performance of a random classifier.}
  \label{fig:S7}
\end{figure}

To test the predictive relevance of the DII critical point, we trained a multilayer perceptron (MLP) binary classifier on the feature subsets selected at each elimination step, for each random subsample, in the unsupervised backward feature elimination via the DII. We test the performance on a held out balanced independent test set (see Section II) computing the area under the receiver operating characteristic (AUROC). Fig. ~\ref{fig:S7} shows the results, analogous to Fig. 4, for the RBP, MEMBRANE and ENZYME datasets. For the physico-chemical features, classification performance saturates at $F \approx F_c$, showing that the DII phase transition coincides with the onset of optimal generalization. In other words, the critical point identified by the DII marks the minimal subset of features retaining full discriminative power (Fig.~\ref{fig:S7}A-C-E). For structural features, the AUROC does not show a plateau increasing F, but it continues increasing. The ENZYME dataset instead shows a saturation of the AUROC increasing $F$, especially for larger $N$, likely due to the fact that it has fewer proteins than the others (see Section II). Additionally, while for LLPS, RBP and MEMBRANE datasets the AUROC values achieved with the full set of features are similar between physico-chemical and structural features, for the ENZYME dataset the structural features achieve higher performance. 

\section*{VII. Software and Reproducibility}
All analyses were implemented in Python 3.9, using DADApy v0.3.2~\cite{glielmo22dadapy} for DII computation, NumPy, scikit-learn, and Matplotlib. Scripts and processed datasets are available at \url{https://github.com/tartaglialabIIT/DIIPhaseTransition.git} and \url{https://doi.org/10.5281/zenodo.18223323}.

\bibliography{biblio}